\def\year{2021}\relax
\documentclass[letterpaper]{article} 
\usepackage{aaai21}  
\usepackage{times}  
\usepackage{helvet} 
\usepackage{courier}  
\usepackage[hyphens]{url}  
\usepackage{graphicx} 
\usepackage{subfig}
\usepackage{natbib}  
\usepackage{caption} 
\usepackage{xcolor}
\usepackage{algorithm} 
\usepackage{algpseudocode} 
\usepackage{amsmath}
\setcitestyle{authoryear}

\newcommand{\revision}[1]{\textcolor{black}{#1}}

\title{Shape-based Feature Engineering for Solar Flare Prediction}
\author{
    Varad Deshmukh\textsuperscript{\rm 1},
    Thomas Berger\textsuperscript{\rm 2},
    James Meiss\textsuperscript{\rm 3}, and
    Elizabeth Bradley\textsuperscript{\rm 1,4}
}
\affiliations{

    \textsuperscript{\rm 1}Department of Computer Science, University of Colorado Boulder, Boulder CO 80309\\
    \textsuperscript{\rm 2}Space Weather Technology Research and Education Center, Boulder CO 80309\\
    \textsuperscript{\rm 3}Department of Applied Mathematics, University of Colorado Boulder, Boulder CO 80309\\
    \textsuperscript{\rm 4}The Santa Fe Institute, Santa Fe NM 87501\\
	$\lbrace$varad.deshmukh, thomas.berger, james.meiss, elizabeth.bradley$\rbrace$@colorado.edu

}

\begin{document}

\maketitle

\begin{abstract}
Solar flares are caused by magnetic eruptions in \textit{active
regions} (ARs) on the surface of the sun.  These events can have
significant impacts on human activity, many of which can be mitigated
with enough advance warning from good forecasts.  To date, machine
learning-based flare-prediction methods have employed physics-based
attributes of the AR images as features; more recently, there has been
some work that uses features deduced automatically by deep learning
methods (such as convolutional neural networks).  We describe a suite
of novel shape-based features extracted from magnetogram images of the
Sun using the tools of computational topology and computational
geometry.  We evaluate these features in the context of a multi-layer
perceptron (MLP) neural network and compare their performance against
the traditional physics-based attributes.  We show that these abstract
shape-based features outperform the features chosen by the human
experts, and that a combination of the two feature sets improves the
forecasting capability even further.
\end{abstract}

\section{Introduction}
\label{sec:introduction}

Solar flares are caused by rearrangement of magnetic field lines in
active regions (ARs) on the surface of the Sun.  These bright flashes
arise from the collision of accelerated charged particles with the
lower solar atmosphere.  The coronal mass ejections (CMEs) that can
accompany these events can have a significant impact on a range of
human activity: damaging spacecraft, creating radiation hazards for
astronauts, interfering with GPS, and causing power grid failures,
among other things.  Lloyd's has estimated that a power outage from an
event associated with a powerful solar flare could produce an economic
cost of 0.6 to 2.6 trillion
dollars \citep{maynard2013}. \revision{Many of these losses could be
mitigated with enough advance accurate warning of impending solar
flares and the accompanying CMEs through actions such as switching to
higher frequency radio for over-the-horizon communications with
international airline flights, preparing satellites in orbit for
safe-mode operations, and bringing additional generation capacity
online to balance power grids against possible geomagnetically induced
current disturbances. Since we currently lack these accurate advanced
warnings, research into how to create them is a high priority.}

\begin{figure*}[htbp]
	\begin{center} \subfloat[]{ \includegraphics[height=0.20\textwidth]{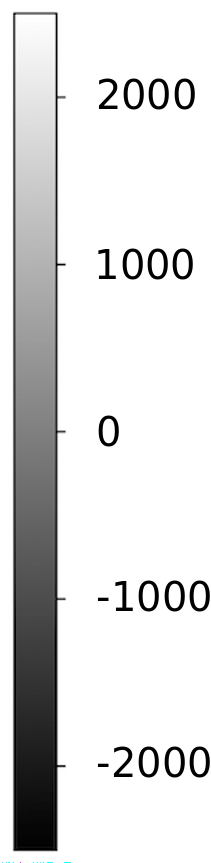}
	    } \subfloat[(a)]{ \includegraphics[height=0.20\textwidth,width=0.20\textwidth]{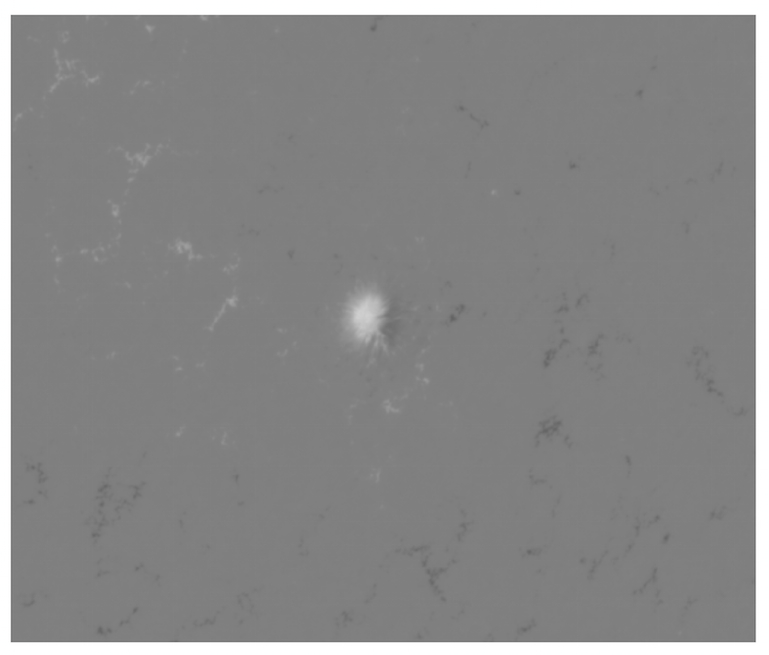}
	    } \subfloat[(b)]{ \includegraphics[height=0.20\textwidth,width=0.20\textwidth,angle=180,
	    origin=c]{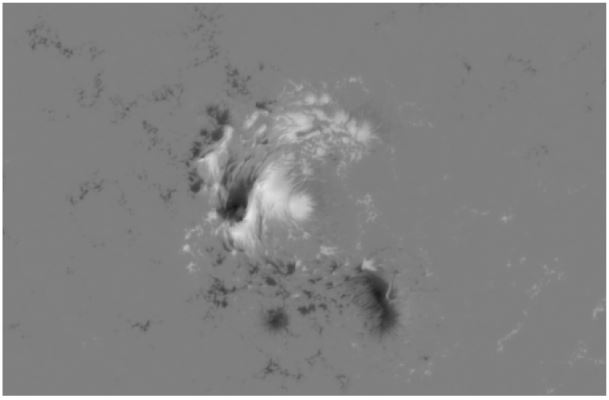}
	    } \subfloat[(c)]{ \includegraphics[height=0.20\textwidth,width=0.20\textwidth,angle=180,
	    origin=c]{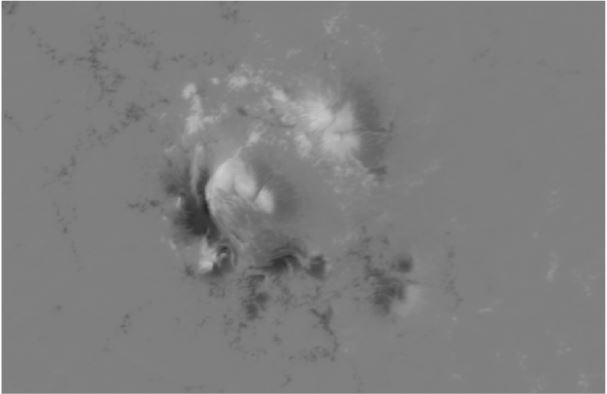}
	    } \end{center} \caption{Three observations of line-of-sight
	    magnetograms of sunspot \#AR 12673, which produced multiple
	    major (M-class and X-class) flares as it crossed the disk of the Sun in
	    September 2017: (a) at 0000 UT on 9/1, (b) at 0900 UT on
	    9/5, about 24 hours before producing an
	    X-class solar flare, and (c) at 1000 UT on 9/7, around the
	    time of an M-class flare.}
	\label{fig:magnetograms}
\end{figure*}

Strategies for flare forecasting rest on the fact that the complexity
of the magnetic field in an AR is known to be relevant to solar-flare
occurrence.  Figure \ref{fig:magnetograms} shows three observations at
different times of the line-of-sight (LOS) magnetic field---called
a \textit{magnetogram}---observed from the sunspot AR 12673 as it
evolved from a simple configuration as seen in panel (a) to more
complex configurations seen in panels (b) and (c).
The white and dark regions represent the LOS magnetic field exiting
and entering the Sun's surface (termed positive and negative polarity,
respectively).  This particular AR produced a powerful flare within 24 hours of
the complex mixed-polarity state observed in panel (b).

It is no surprise that these kinds of magnetic field observations have
played a central role in machine learning-based forecasting models for
solar flares.  Typically, this has involved the use of features that
solar-physics experts consider to be revelant to solar flaring, such
as the magnetic field or electric current strength, current helicity,
magnetic shear, and the like.\footnote{Please refer to Table 1
of \citet{Deshmukh2020} for a complete list and to \citet{Bobra:2014dn}
for details about the associated calculations.}  
Recently, there has been a push to use convolutional neural networks
(CNNs) to automatically learn latent features that are statistically
correlated to the occurence of a solar flare.  In this work, we take a
wholly different approach, defining a novel feature set based purely
on the shapes of the structures in the magnetogram.  We formally
quantify the complexity of an active region by using
\textit{computational geometry} and \textit{computational
topology} techniques on the radial component of the photospheric
magnetic field, focusing specifically on the proximity and interaction
of the polarities, as well as the components and holes in sub-level
thresholded versions of the magnetogram image.  Following a brief
review of ML-based flare forecasting work and a description of the
data, we present the results of a comparative study about the efficacy
of these features in a multi-layer perceptron model.

In operational space weather forecasting offices, human forecasters
currently use the McIntosh \citep{McIntosh:1990wu} or
Hale \citep{Hale:1919} classification systems to categorize active
regions into various classes; they then determine the statistical
24-hour flaring probability derived from historical
records \citep{Crown:2012}.
Over the past decade, significant effort has been devoted to
machine-learning solutions to this problem, including support vector
machines (SVM)
\citep{Bobra:2015fn,boucheron2015,Nishizuka:2017,yang2013,yuan2010},
multi-layer perceptron (MLP) models \citep{Nishizuka:2018}, Bayesian
networks
\citep{yu2010}, logistic
regression \citep{yuan2010}, LASSO regression \citep{Campi2019},
linear classifiers \citep{Jonas:2018}, fuzzy
C-means \citep{Benvenuto2018} and random
forests \citep{Campi2019, Nishizuka:2017}.
Recently, the ML-based flare forecasting community has turned to deep
learning methods for automatically extracting important features from
raw image data that are relevant for flare-based
classification \cite{Chen2019,Huang2018, Park2018, Zheng2019}.  The
work cited in this paragraph is only a representative subset of
ongoing research in this active field; for a more complete
bibliography, please refer to \citet{Deshmukh2020}.

In this paper, we use magnetograms from the Helioseismic and Magnetic
Imager (HMI) instrument onboard NASA's Solar Dynamics Observatory
(SDO), which has been deployed since 2010.  Rectangular cutouts of
each AR on the disk of the sun in each of these images, termed
Spaceweather HMI Active Region Patches (SHARPs)---three examples of
which make up Figure \ref{fig:magnetograms}---are available to download
from the Joint Space Operations Center website ({\tt \small
jsoc.stanford.edu/}).  The metadata that accompanies each SHARP record
contains values for the physics-based features mentioned above: i.e.,
the attributes that domain experts consider meaningful for the physics
of the system.  The dataset for the study reported in this paper,
which covers the period from 2010-2016 at a one-hour cadence, focuses
specifically on the radial magnetic field component from these
images because of its role in magnetic reconnection.

The active regions in this dataset---which contains about 2.6 million
data records, each approximately 2 MB in size, totaling 5 TB of
data---are known to have produced about $1250$ major flares within 24
hours of the image time \citep{Schrijver:2016cs}.  We use the NOAA
Geostationary Operational Environment Satellite (GOES) X-ray
Spectrometer (XRS) flare catalog to identify these events and label
the associated SHARP with a 1 if it produced a major flare---one whose
peak flux in the 1-8 \AA\ range is greater than $10^{-5} W/m^2$---in
the 24 hours following the time of the sample, and 0 otherwise.  Next,
we discard all the magnetogram images that contain invalid pixel data
(NaN values).  The resulting data set included $3691$ active regions,
of which $141$ produced at least one major flare as they crossed the
Sun's disk and $3550$ did not.  This corresponded to $438,539$ total
magnetograms, of which $5538$ and $432821$, respectively, were labeled
as flaring and non-flaring.

A large positive/negative imbalance like this (78:1) is an obvious
challenge in a binary classification machine-learning problem, as described at more length
below.  Another issue is that multiple images are available from a
single AR during the run-up to a particular flare.  To avoid
artificially boosting our model accuracy by, for example, testing on
an image that is one hour earlier than, and thus very similar to, an
image in the training set, we perform an additional check each time we
split the data into training (70\%) and testing (30\%) sets to ensure
that all the magnetogram images belonging to a given AR are grouped
together and placed either in the training or the testing set. 10 different
random seeds are used for shuffling the data to generate 10 training/testing 
set combinations.

\section{Shape-based Featurization of Active Regions}

As in many machine-learning problems, the choice of features is
critical here.  Quantitative comparison studies show that none of the
methods described above that use physics-based features extracted from
magnetic field data are significantly more skilled---and indeed are
typically less skilled---than current human-in-the-loop operational
forecasts \citep{Barnes:2016bu,leka2019a,leka2019b}.  In other words,
while the physics-based attributes are no doubt important, they may
not necessarily form an effective feature set for solar-flare
forecasting.

The novelty of our work is our approach to the feature-engineering
task from a mathematical standpoint, rather than a physics-based one.
Specifically, we use computational topology and computational geometry
to extract features that are based purely on the shapes of the regions
in the magnetograms.  The underlying conjecture is that this is a
useful way to capture the complexity of these regions---which is known
to be related to flaring.  As preliminary evidence in favor of that
conjecture, we show that shape-based features outperform the
traditional physics-based features in the context of a multi-layer
perceptron model, yielding a better 24-hour prediction accuracy.

Note that our objective in this work is not to directly compare our
forecasting model with other methods, but to primarily convince the
reader of the importance of shape-based features for solar flare
forecasting.

\subsection{Computational Geometry}

To compute geometry-based features from each magnetogram, we first
remove noise by filtering out pixels whose magnetic flux magnitude is
below $200\ G$, then aggregate the resulting pixels into clusters if
they touch along any side or corner.  We then determine the number and
area of each cluster, discarding all whose area is less than 10\% of
the maximum cluster area.  We perform these operations separately for
the positive ($>200\ G$) and negative ($<-200\ G$) fields.  

\begin{figure}[h]
	\begin{center} \subfloat[\revision{(a)}]{ \includegraphics[width=0.227\textwidth]{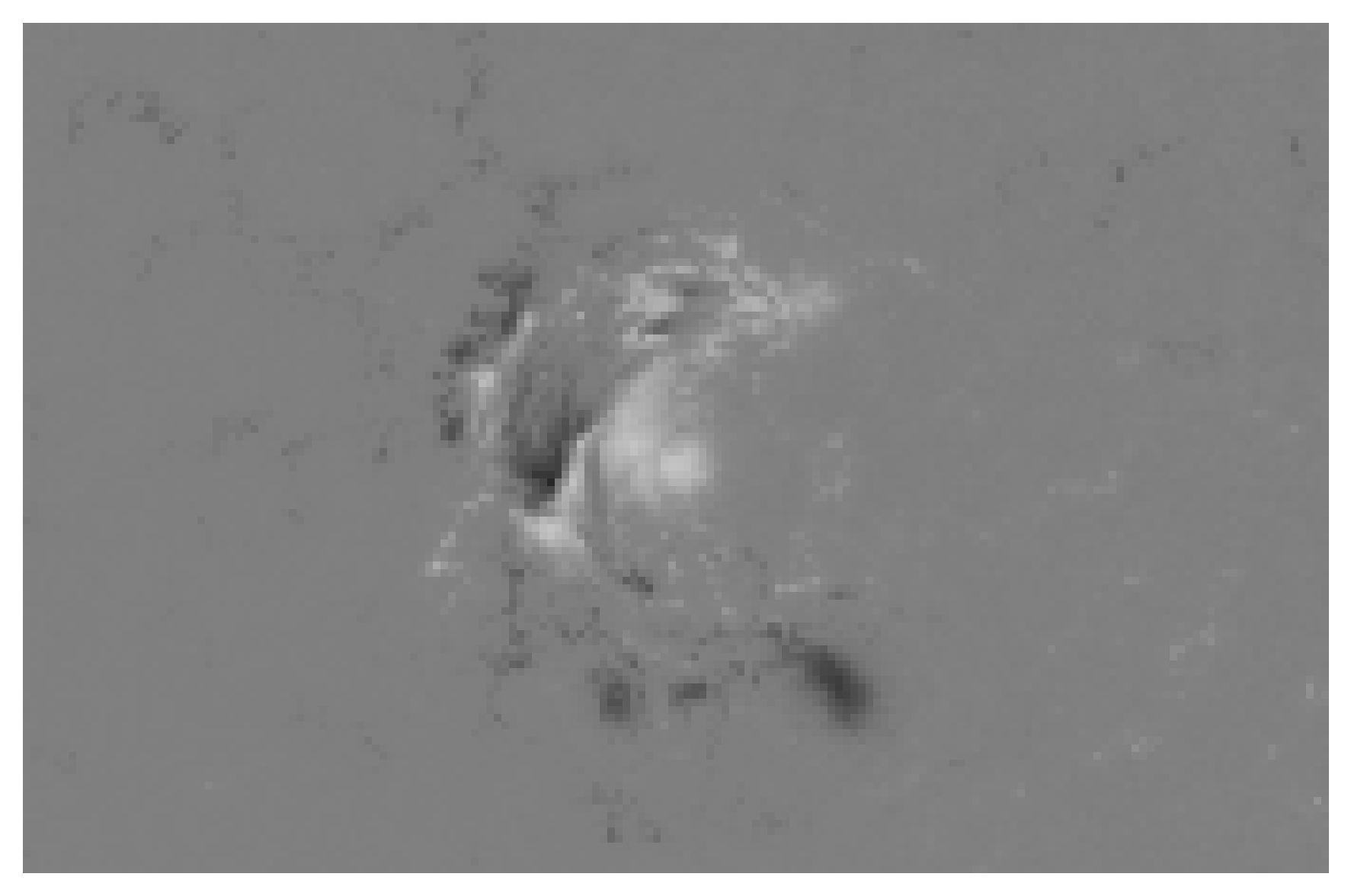}}\\ \subfloat[\revision{(b)}]{ \includegraphics[width=0.227\textwidth]{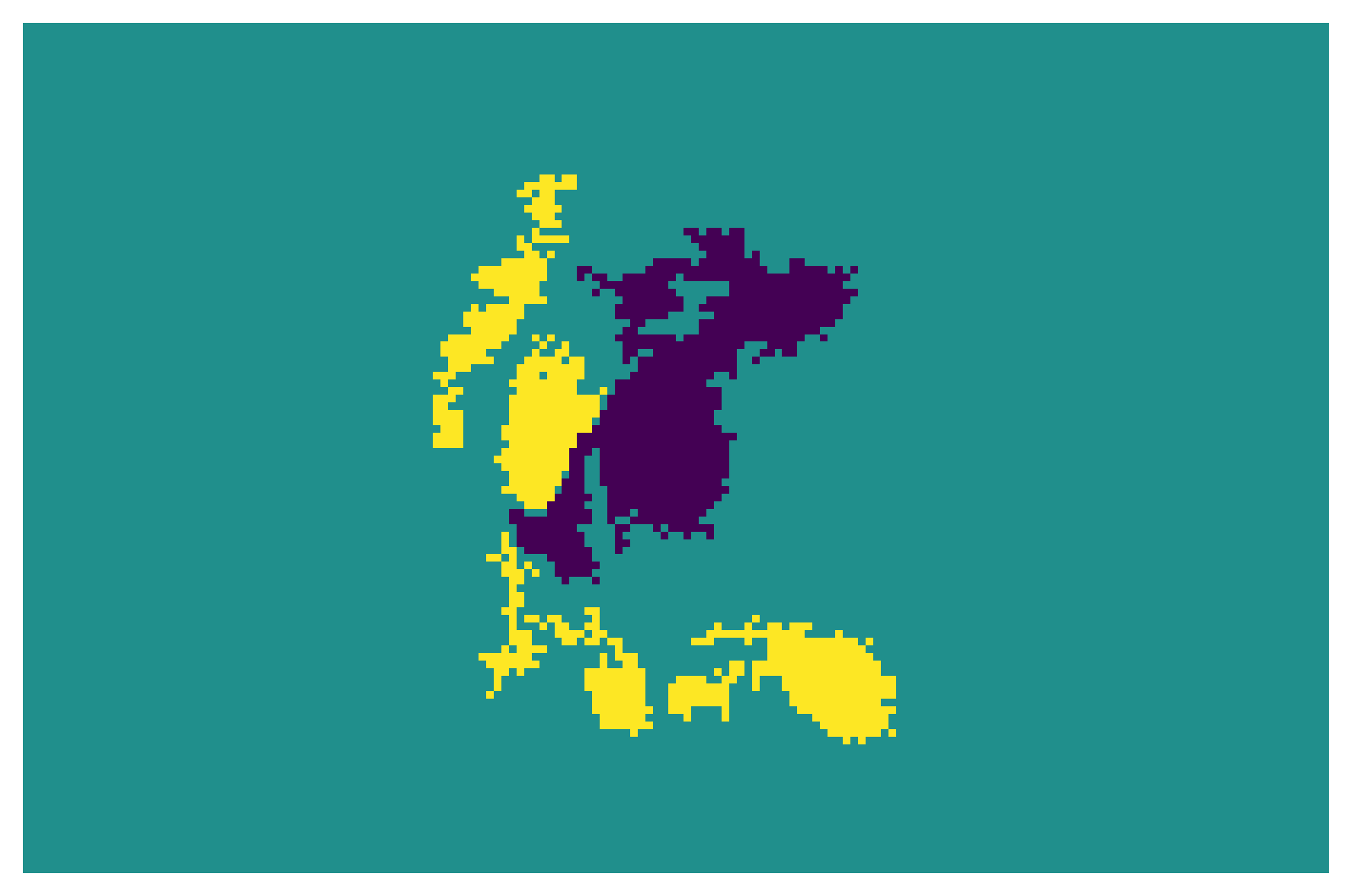}} \subfloat[\revision{(c)}]{ \includegraphics[width=0.227\textwidth]{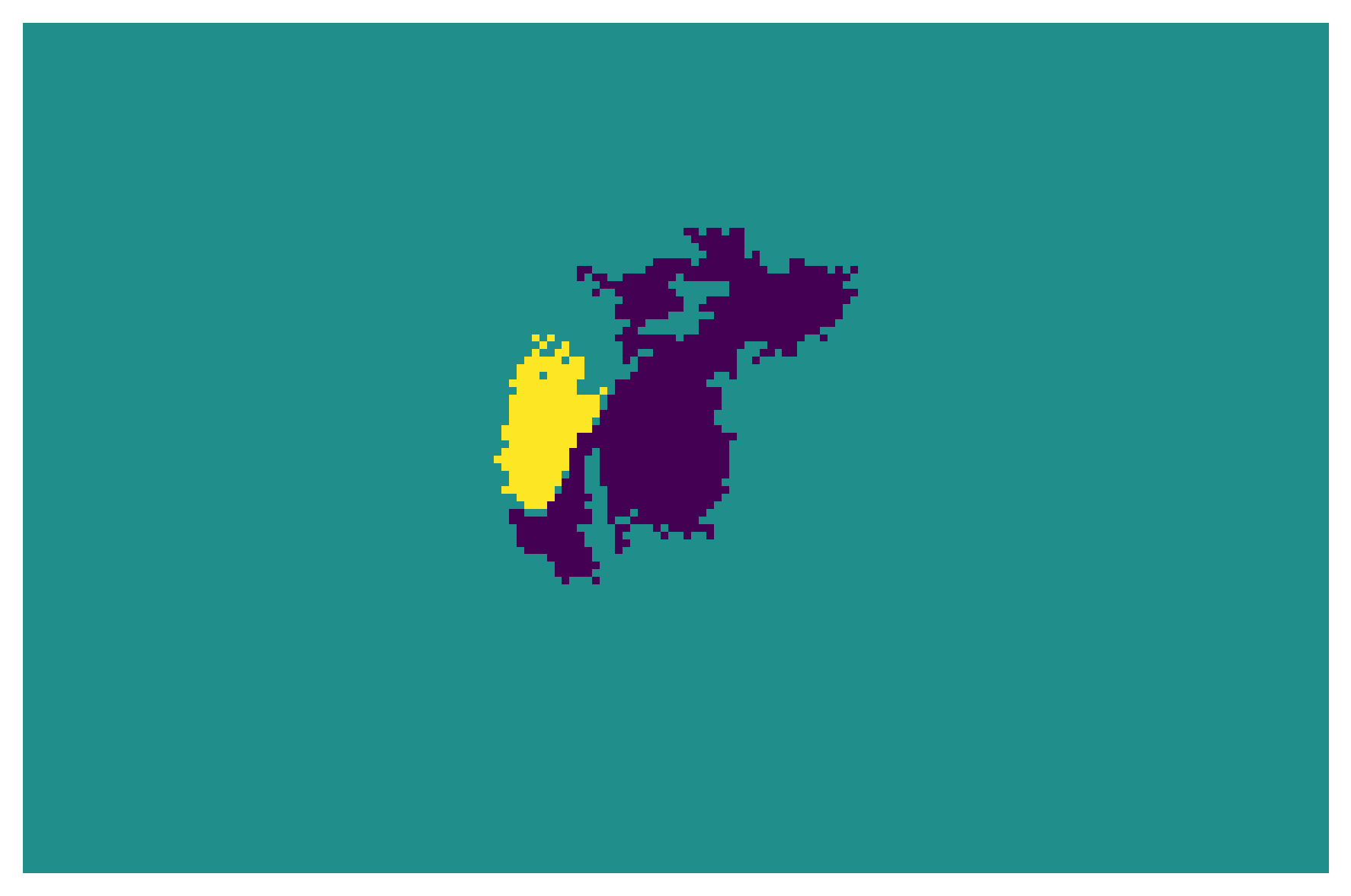}} \end{center} \caption{\revision{Process
		for determining the most interacting postive/negative
		cluster pair in geometry-based feature extraction.
		From a sample magnetogram shown in panel (a), positive
		(blue) and negative (yellow) clusters of a sufficiently large 
		size are extracted (panel b); from these, the most interacting 
		cluster pair is determined via calculations of the magnetic
		flux in each of the paired regions (panel
		c).}}  \label{fig:geometry_features}
\end{figure}

We then compute an {\sl interaction factor (IF)} between all
positive/negative polarity pairs, defined in a manner similar to the
so-called Ising Energy used by \citet{Florios2018} (introduced first
in \citeauthor{Ahmed2010}, \citeyear{Ahmed2010}):
%
%
\begin{equation}
	IF=\frac{B_{pos} \times B_{neg}}{r_{min}^2}
	\label{eqn:interaction}
\end{equation}
\noindent where $B_{pos}$ and $B_{neg}$ are the sums of the flux over the
respective components and $r_{min}$ is the smallest distance between
them.  A high $IF$ value is an indication of strong, opposite-polarity
regions in close proximity---an ideal configuration for a flare.
Following this reasoning, we choose the pair with the highest $IF$
value and derive a number of secondary features from it, such as the
center of mass distance between the two clusters. Extraction of the
most interacting pair on an example magnetogram is shown in
Figure~\ref{fig:geometry_features}.  Together with the values used in
the computation of $IF$---the magnetic flux of the positive and
negative clusters, the center of mass distance between them, the
smallest distance between them, the interaction factor, etc.---these
make up the 16-element feature vector that quantifies the interaction
of the opposite polarity regions. The feature extraction process 
together with the final list of geometry-based features is summarized in 
Algorithm~\ref{alg:geometry}.\footnote{Please refer to Table 2 of 
\citet{Deshmukh2020} for a 
complete description.}

\begin{algorithm*}[h]
	\caption{\revision{Geometry-based Feature Extraction}}
	\begin{algorithmic}[1]
		\For {\revision{each SHARPs magnetogram image}}
		\State \revision{Cap magnitude of all pixels to $200G$ from below, preserving the sign of each pixel.}
		\State \revision{Find positive and negative flux clusters in the magnetogram.}
		\State \revision{Remove clusters with area less than 10\% of the maximum cluster size.}
		\For {\revision{each pair of positive and negative clusters $\{B_{pos}, B_{neg}\}$}}
		\State \revision{Compute the interaction factor IF (Eqn.~\ref{eqn:interaction}).}
		\EndFor
		\State \revision{Determine the pair with the maximum IF; call it the most interacting pair (MIP): $\{B_{pos}, B_{neg}\}^{max}$.}
		\State \revision{Extract 16 geometry-based features: total positive and negative clusters in the magnetogram (2), areas of the largest positive and negative cluster (2), total magnetic fluxes of the largest positive and negative cluster (2), IF (1), MIP center of mass distance (1), MIP smallest distance (1), ratio of the MIP center of mass distance to the MIP smallest distance (1), total magnetic fluxes of the MIP clusters (2), areas of the MIP clusters (2) and total flux densities of the MIP clusters (2).}
		\EndFor
	\end{algorithmic} 
	\label{alg:geometry}
\end{algorithm*} 

\begin{algorithm*}[h]
	\caption{\revision{Topology-based Feature Extraction}}
	\begin{algorithmic}[1]
		\For {\revision{each SHARPs magnetogram image}}
		\State \revision{Compute  $\beta_1$ persistence diagrams using a cubical complex algorithm
		          for positive and negative flux values.}
		\State \revision{Count the number of ``live" $\beta_1$ holes for 20 flux values in the range $[-5000G, 5000G]$.}
		\EndFor
	\end{algorithmic} 
	\label{alg:topology}
\end{algorithm*} 

\subsection{Computational Topology}

Computational topology, also known as topological data analysis
(TDA) \citep{Ghrist08,Kaczynski04,Zomorodian12}, operationalizes the
abstract mathematical theory of shape to allow its use with real-world
data.  These methods, which have been used to advantage in
applications ranging from biological aggregation models \citep{topaz}
to the large-scale structure of the universe \citep{Xu:2018xnz},
provide a useful strategy for extracting and codifying the spatial
richness of magnetograms like the ones shown in
Figure \ref{fig:magnetograms}.

The {\sl homology} of an object formally quantifies its shape using
the Betti numbers: the number of components ($\beta_0$), holes
($\beta_1$), voids ($\beta_2$), and so on.  When one has a smooth,
well-defined object, the textbook formulation of homology addresses
this quantification, but real-world data---a finite collection of
points or a set of pixels---does not really have a ``shape.''  TDA
handles this by filling in the gaps between the data points with
different types of simplices.  The simplest way to do this maps well
to pixellated images; one can create a manifold from a selected set of 
pixels in an image by replacing each one by a cubical simplex---a
square piece complete with its vertices and edges.  This leads to the
notion of connectedness amongst discrete pixels: a pair of pixels are
said to be ``connected'' if their corresponding cubical simplices
share an edge or a vertex.  Such connections lead to the formation of
different connected components, holes, etc.

\begin{figure}[H]
	\begin{center} \subfloat[(a) Example
	    Image]{ \includegraphics[width=0.15\textwidth]{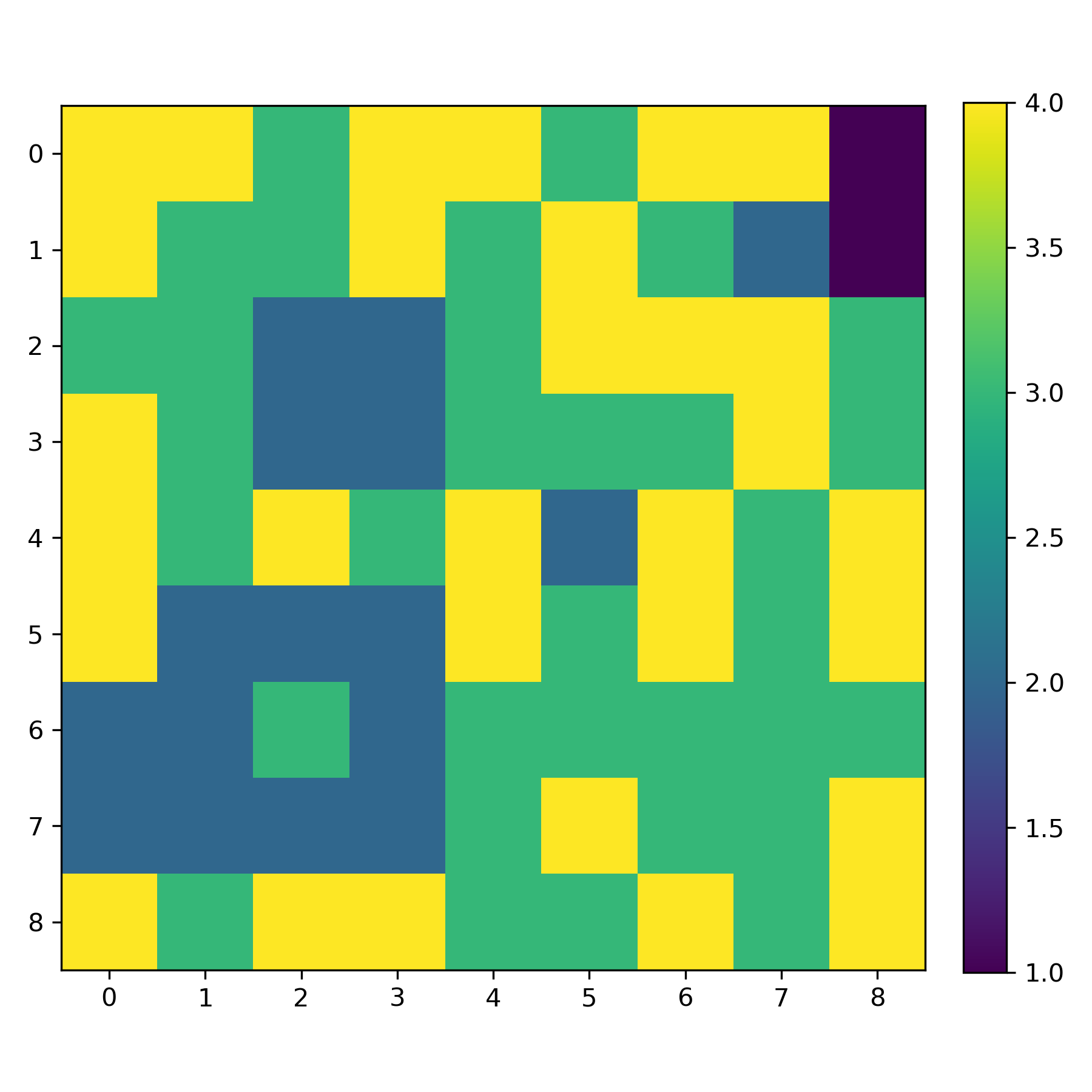}
	    } \subfloat[(b) $t$=0]{ \includegraphics[width=0.15\textwidth]{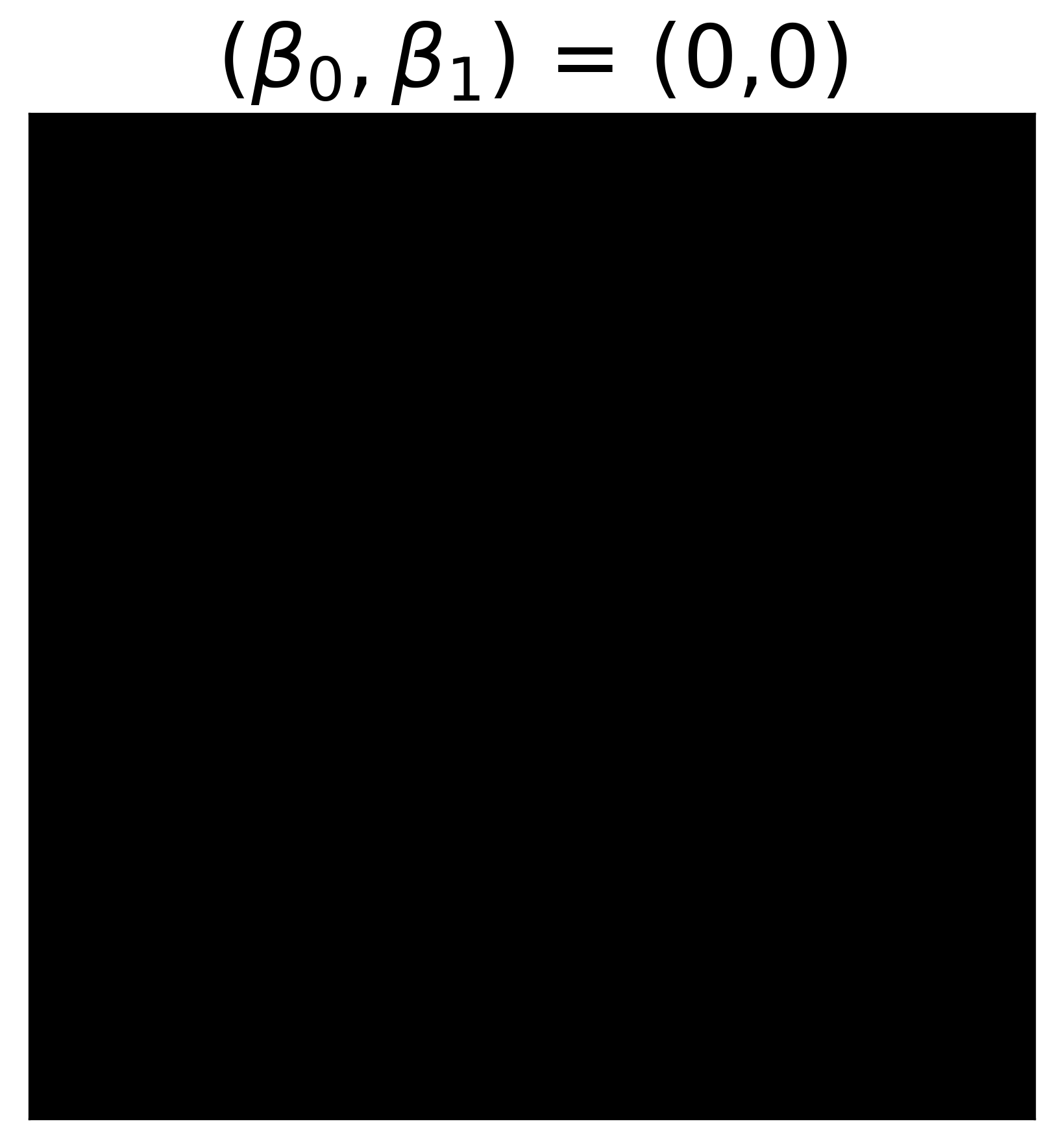}
	    } \subfloat[(c) $t$=1]{ \includegraphics[width=0.15\textwidth]{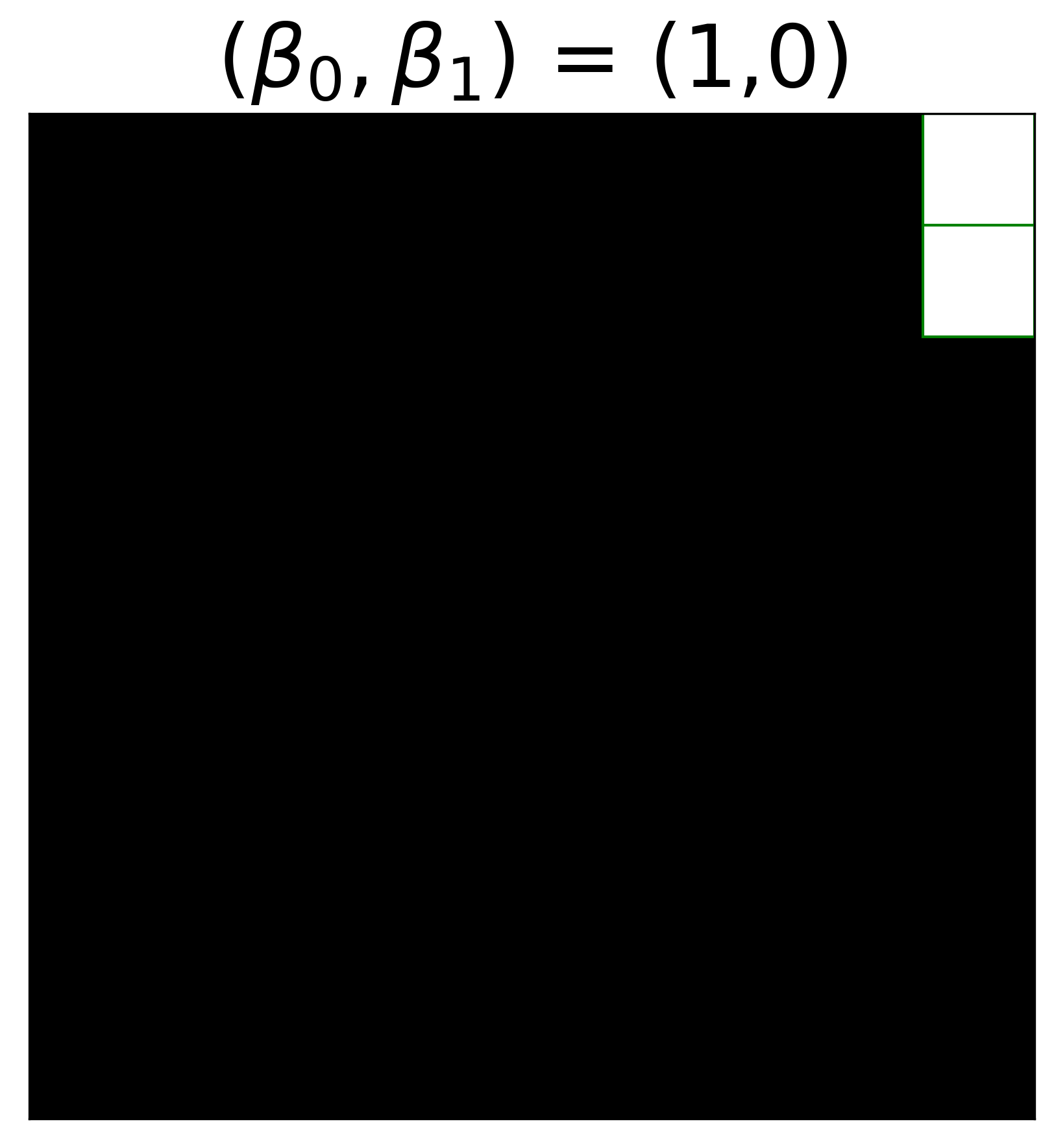}
	    }\\ \subfloat[(d) $t$=2]{ \includegraphics[width=0.15\textwidth]{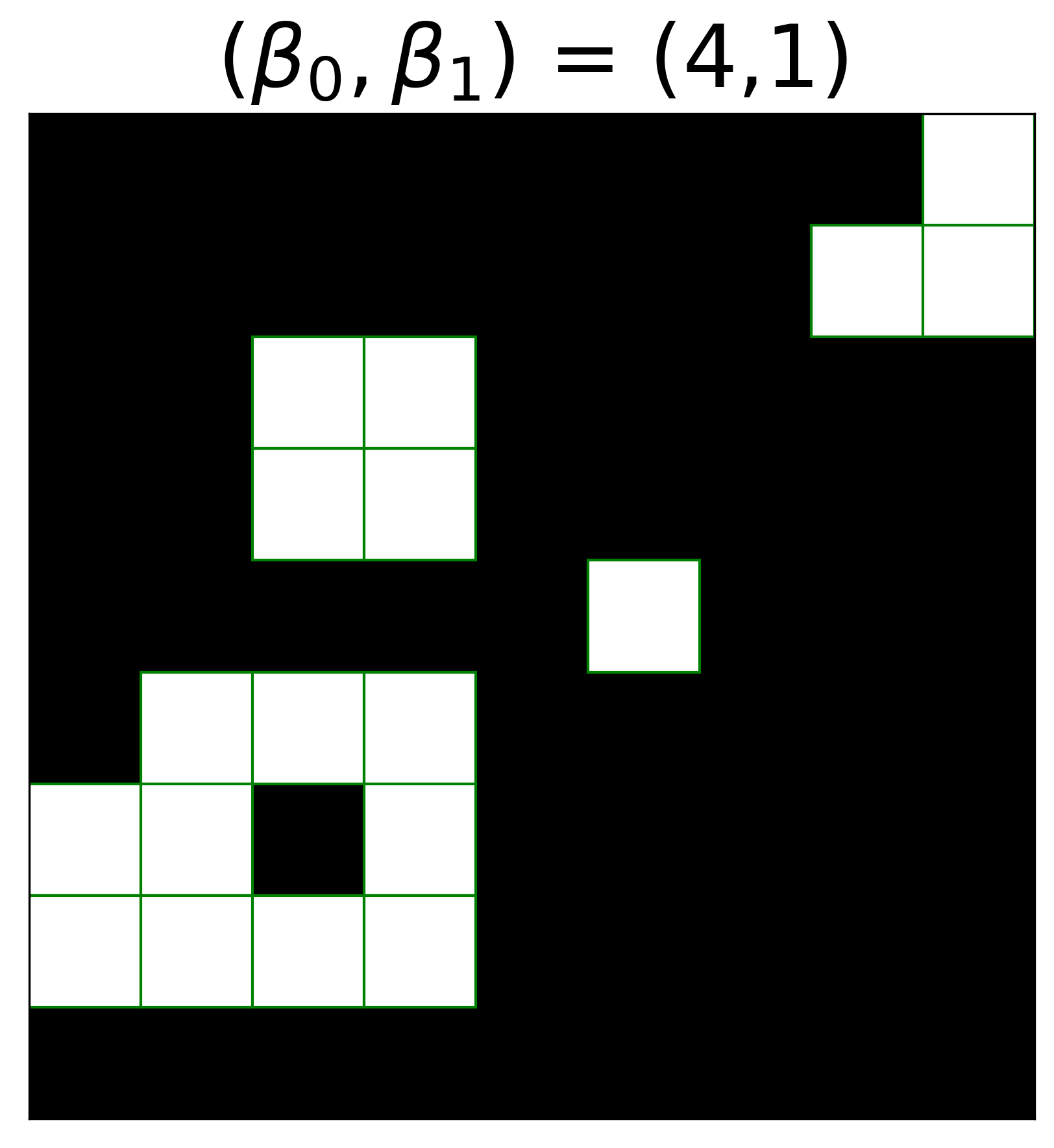}
	    } \subfloat[(e) $t$=3]{ \includegraphics[width=0.15\textwidth]{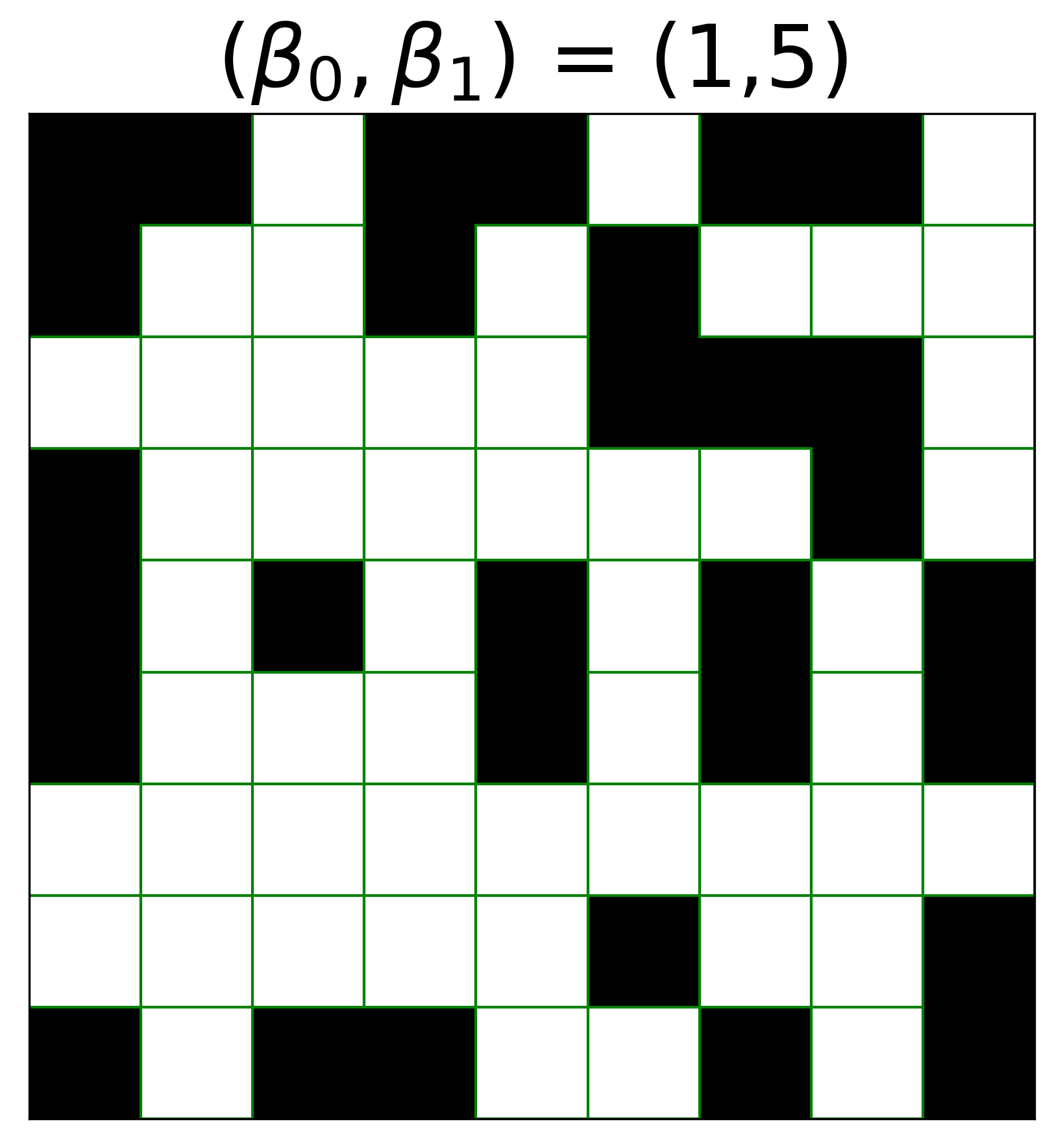}
	    } \subfloat[(f) $t$=4]{ \includegraphics[width=0.15\textwidth]{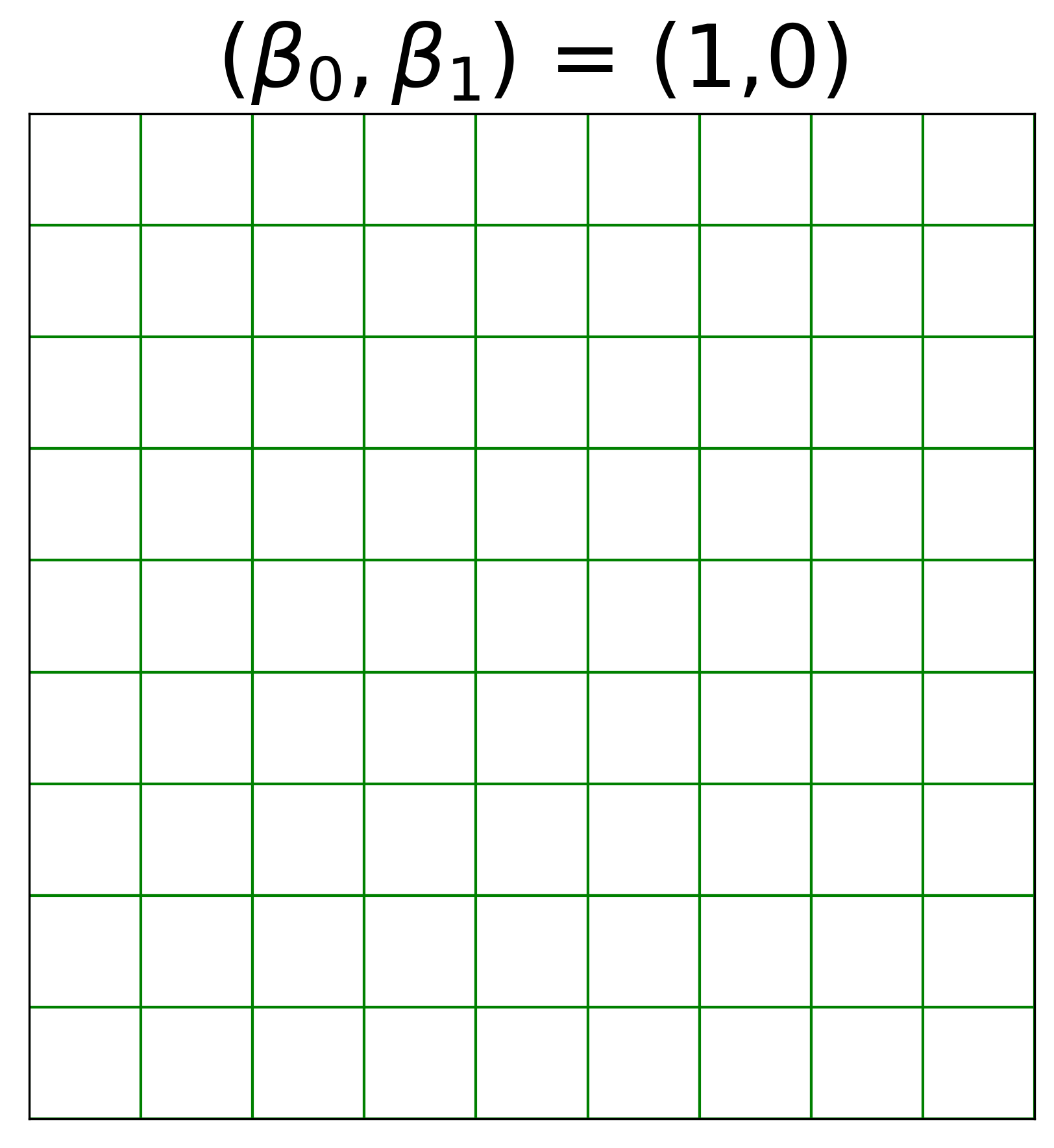}
	    }\\ \subfloat[(g) $\beta_1$ Persistence
	    Diagram]{ \includegraphics[width=0.20\textwidth]{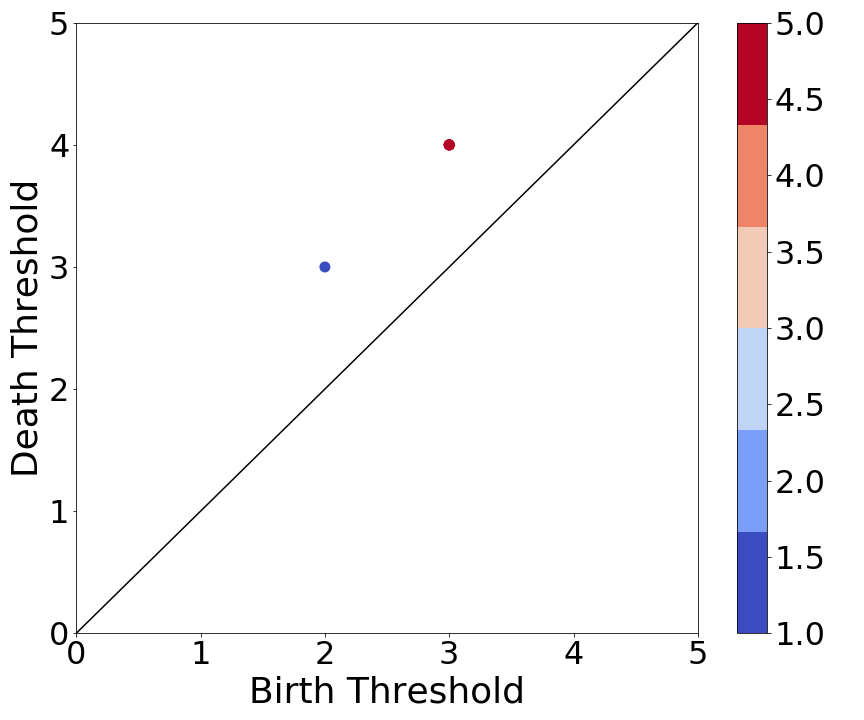}
	    } \end{center} \caption{Computational topology: (a)
	    Image-based dataset. (b)-(f) Cubical complex of that
	    dataset for five values of sub-level thresholding
	    ($t=[0,1,2,3,4]$). For each complex, the threshold $t$ and
	    the $(\beta_0,\beta_1)$ counts are mentioned. (g):
	    $\beta_1$ Persistence diagram.} 
    \label{fig:components}
\end{figure}

In images where the pixel values range over some interval, it can be
useful to combine this idea with thresholding.
Figure \ref{fig:components} demonstrates the process of generating a
cubical complex for a range of threshold values $t$.
When the threshold is low, as in Figure \ref{fig:components}(b), none
of the pixels are in the complex ($\beta_0=0$) and it has no holes
($\beta_1=0$).  As $t$ is raised and lower-value pixels enter the
computation, the complex develops a small connected component at the
top right ($\beta_0=1$).  Four different components can be observed in
Figure \ref{fig:components}(d) for a threshold $t = 2$; at $t=3$, all
the components become merged together.  In addition to the formation
of components, two-dimensional ``holes'' are also formed when edges
from various cubical simplices form a loop in the complex that is not
filled by a cubical simplex (dark regions surrounded by green edges on
all sides). We can see the presence of one and five holes, respectively,
for $t = 2$ and $t = 3$.

This formation and merging of the various components and holes with
changing threshold captures the shape of the set in a very nuanced
way.  The idea of {\sl persistence}, first introduced
in \citet{Edelsbrunner00} 
(and independently by \citeauthor{Robins02}, \citeyear{Robins02}), 
is that tracking that evolution allows one to deduce important
information about the underlying shape that is sampled by these
points.  To capture all of this rich information, one can use a single
plot called a {\sl persistence diagram}~\citep{Edelsbrunner00}.  Most
components, for example, have birth and death parameter values, where
they appear and disappear, respectively, from the construction.  A
$\beta_0$-persistence diagram has a point at ($t_{birth}$,
$t_{death}$) for each component, while a $\beta_1$-persistence diagram
(PD) does the same for all the holes.  The $\beta_1$ PD for our toy
image example is shown in Figure \ref{fig:components}(g).  Multiplicity
of different holes with the same ($t_{birth}$, $t_{death}$) is
represented by color; the single hole that formed at $t=2$ and died at
$t=3$ is represented in blue, whereas the five holes corresponding to
(3,4) are colored red.

\begin{figure*}[h]
	\begin{center}
		\subfloat[0000 UT on 9/1/17] {
			\includegraphics[height=0.20\textwidth]{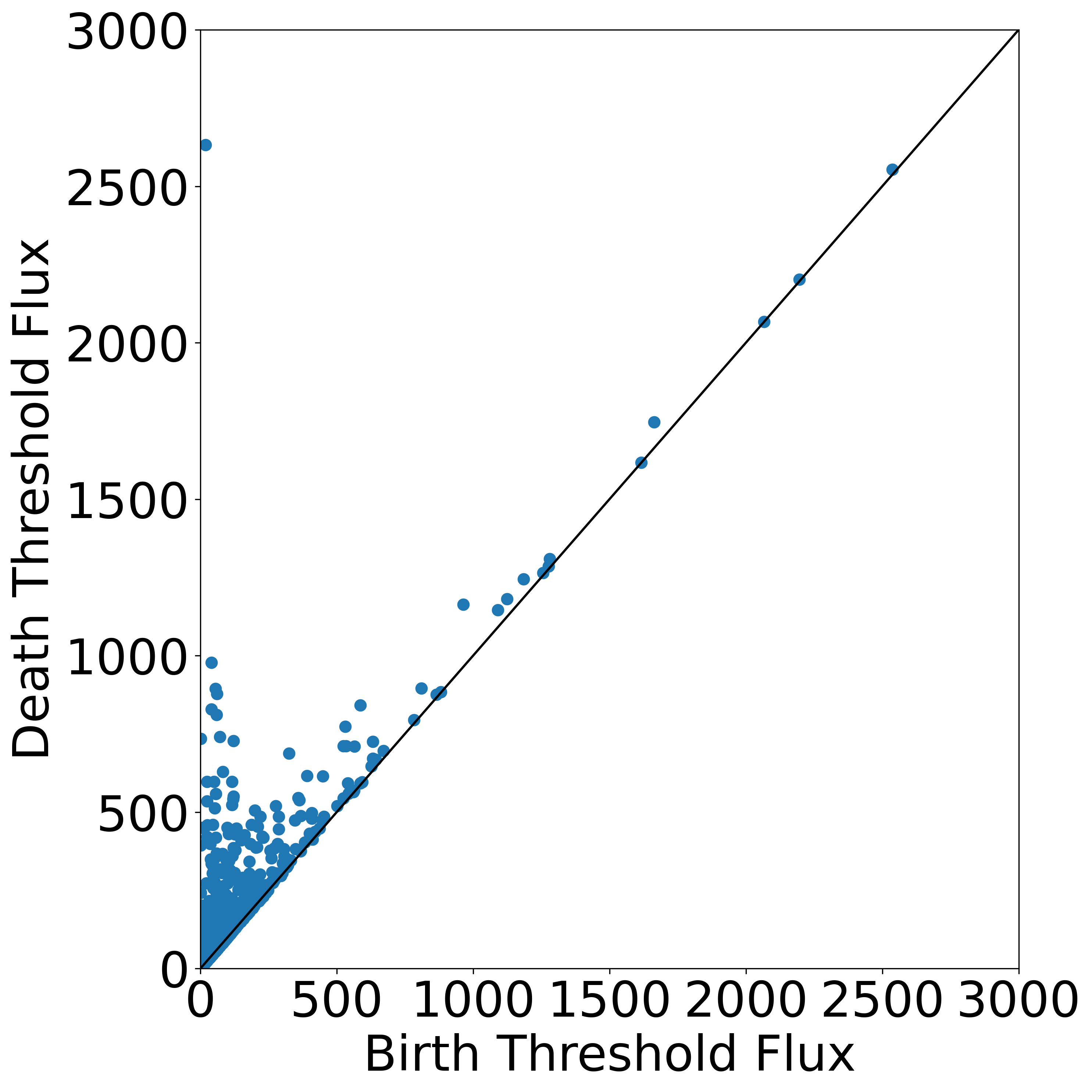}
		}
		\subfloat[0900 UT on 9/5/17] {
			\includegraphics[height=0.20\textwidth]{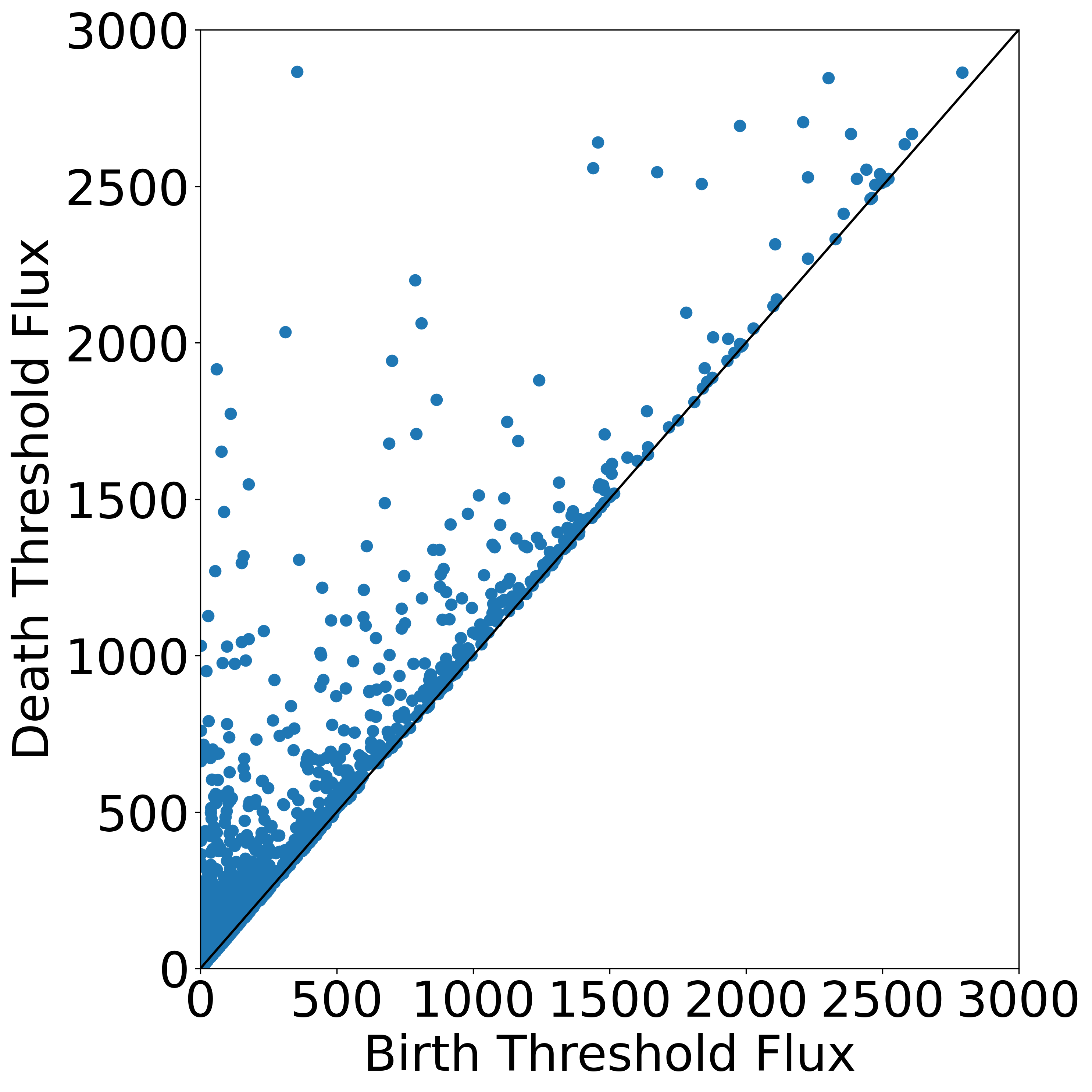}
		}
		\subfloat[1000 UT on 9/7/17] {
			\includegraphics[height=0.20\textwidth]{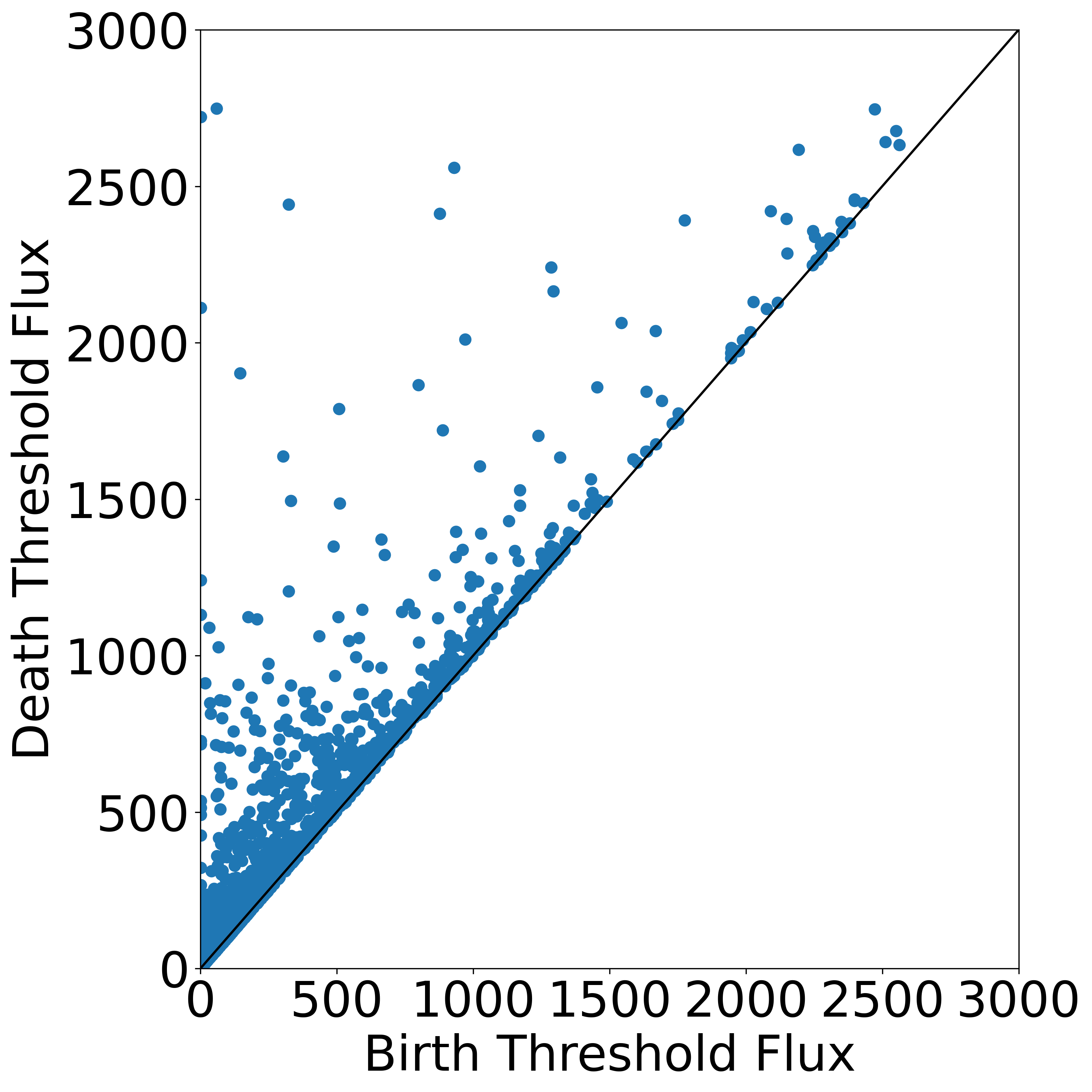}
	        }	
	\end{center}
%
%
	\caption{$\beta_1$ persistence diagrams for the magnetograms
		of Figure \ref{fig:magnetograms}, constructed from the
		set of pixels with positive magnetic flux
		densities using the cubical complex approach.  These
		diagrams reveal a clear change in the topology of the
		field structure well before the major flare that
		was generated by this active region at 0910 UT on 6
		September 2017.}  \label{fig:pds}
\end{figure*}

The $\beta_1$ persistence diagram is the basis for our topology-based
feature set.  For each magnetogram, we first generate separate PDs for the
positive and negative polarities.  Figure \ref{fig:pds} shows $\beta_1$
PDs for the positive flux field in the series of magnetograms in
Figure \ref{fig:magnetograms}.
The increase in the complexity of the AR between 2017-09-01 00:00:00 UT 
and 2017-09-05 09:00:00 UT
is reflected in the patterns in the PDs: Figure \ref{fig:pds}(b) (24
hours prior to a flare) contains a far larger number of off-diagonal
holes---i.e., those that persist for larger ranges of $t$---than
Figure \ref{fig:pds}(a), which is a newly formed AR.

This visual evidence supports our claim that PDs can effectively
quantify the growing complexity of a magnetogram during the lead-up to
a flare.  The next step is to determine whether that observation
translates to discriminative power in the context of a
machine-learning method.  This requires one more step: vectorization
of the persistence diagrams into a set of features.  For this, we use
a very simple method, choosing a set of 20 flux values in the interval 
$[-5000G, 5000G]$, and counting the number of holes that are ``live" in the PDs
at each of these flux values. Repeating this operation
separately for the positive and negative polarities, we obtain 20
entries for our topology-based feature set. The feature extraction
process is briefly summarized in Algorithm~\ref{alg:topology}.

While our persistence diagram vectorization approach is relatively simple,
there has been a significant effort over the last few years to more efficiently 
vectorize persistence diagrams for using them with ML models 
\citep{Adams2017, Bubenik2015, Carriere2019, Carriere2017, Kusano2016}. 
We plan to incorporate some of these techniques in future work 
to improve our solar flare prediction model.

\section{Machine Learning Model}

As a testbed for evaluating the different feature sets, we design a
standard feedforward neural network using \textsc{Pytorch} with six
densely connected layers.  The input layer size is variable depending
on the size of the feature set; the output layer contains two neurons
corresponding to the two classes---flaring and non-flaring.  The four
intermediate layers contain 36, 24, 16 and 8 neurons respectively,
when counting from the direction of the input to the output layer.  To
prevent over-fitting, a Ridge Regression regularization with a penalty
factor is used at each layer that limits the $L_{2}$ sum of all the
weights. At each hidden layer, a {\sl ReLU} activation is used, with a
{\sl softmax} activation applied to the final layer.  We use an
Adagrad optimizer for updating the model weights during the back
propagation.  A batch size of 128 is used in the gradient descent. The
loss function used for optimization is a weighted binary cross-entropy
error; since the dataset is imbalanced, a weight greater than 1 is
associated with the flaring class to penalize a flare misprediction
more than a non-flare misprediction. Finally, the model is trained over
15 epochs before evaluation.

\subsection{Hyperparameter Tuning}

For each feature set combination, we tune a number of important model
hyperparameters--- the learning rate, the $L_{2}$ penalty regularization
factor, the cross-entropy weight ratio and the learning rate
decay---to ensure that the model is optimized for the corresponding
feature set and the comparison is fair.  Our tuning algorithm is as
follows:
\begin{enumerate}
\item Select 40 different hyperparameter combinations using 
the {\tt python} {\tt bayesopt} library \citep{Ruben2014}, which employs a Gaussian
process-based Bayesian sampling approach.
\item Use a five-fold cross-validation approach to determine the performance 
of each hyperparameter combination by evaluating the average
validation True Skill Statistic (TSS) metric
score \citep{Woodcock1976} across the five folds.
\item Select the hyperparameter combination with the highest 
score and use it to train the model on the full training set, then
evaluate this model on the test set.
\end{enumerate}

This procedure is followed for all 10 training set/testing set splits
of the magnetogram data described earlier.  We use the {\tt ray.tune}
library \citep{tune2018} to parallelize the effort of this computationally intensive
task.  With this setup, each tuning experiment for a single
training-test combination and a single feature set takes about 5 hours
on an NVIDIA Titan RTX GPU.

\section{Results}

To determine whether these geometry- and topology-based feature sets
improve upon, or synergize with, the commonly used physics-based
SHARPs feature sets described in the third paragraph of the introduction, 
we follow the procedure described in the previous section for each
feature set in isolation, as well as in various combinations with the
other sets.

\begin{table*}[htbp]
    \centering
    \begin{tabular}{c c c c c c c}
        \hline
        & Accuracy & Precision & Recall & FB & TSS & HSS \\
        \hline
        \hline
        
        Perfect score & 1 & 1 & 1 & 1 & 1 & 1 \\
        
        \hline \\ 
        SHARPs (19) & 0.84 $\pm$ 0.02 & 0.06 $\pm$ 0.01 & 0.87 $\pm$ 0.05 & 13.84 $\pm$ 1.93 & 0.70 $\pm$ 0.01 & 0.09 $\pm$ 0.02 \\ \\
        
        Geometry (16) & 0.82 $\pm$ 0.01 & 0.06 $\pm$ 0.01 & 0.89 $\pm$ 0.04 & 14.89 $\pm$ 1.15 & 0.71 $\pm$ 0.04 & 0.09 $\pm$ 0.01 \\ \\
        
        Topology (20) & 0.86 $\pm$ 0.02 & 0.08 $\pm$ 0.01 & \textbf{0.90 $\pm$ 0.02} & 12.20 $\pm$ 1.96 & \textbf{0.75 $\pm$ 0.03} & 0.12 $\pm$ 0.02 \\ \\
        
        SHARPs + Geometry (35) & 0.84 $\pm$ 0.02 & 0.07 $\pm$ 0.01 & 0.89 $\pm$ 0.05 & 13.24 $\pm$ 1.98 & 0.73 $\pm$ 0.03 & 0.11 $\pm$ 0.01 \\ \\
        
        SHARPs + Topology (39) & \textbf{0.86 $\pm$ 0.01} & \textbf{0.08 $\pm$ 0.01} & 0.89 $\pm$ 0.03 & \textbf{11.55 $\pm$ 1.06} & \textbf{0.75 $\pm$ 0.03} & \textbf{0.12 $\pm$ 0.01} \\ \\
        
        All three sets (55) & \textbf{0.86 $\pm$ 0.01} & \textbf{0.08 $\pm$ 0.01} & 0.87 $\pm$ 0.04 & 11.77 $\pm$ 1.27 & 0.74 $\pm$
        0.03 & 0.11 $\pm$
        0.01 \\ \hline \end{tabular} \caption{Performance of the
        various feature sets. Numbers in paranthesis indicate 
        the number of elements in the input feature vector. For all 
        the metrics except for frequency bias (FB), higher is better.}  
    \label{tab:results}
\end{table*}
To evaluate the results, we employ a number of standard metrics from
the prediction literature: accuracy, precision, recall, True Skill
Statistic (TSS), Heidke Skill Score (HSS), and frequency bias (FB).
These metrics, which assess correctness in different ways, are derived
from the entries of the contingency table generated by comparing the
model forecast against the ground truth---True Positives (TP), False
Positives (FP), False Negatives (FN) and True Negatives (TN). A description
of these metrics can be found in \citet{Crown:2012} and \citet{leka2019a}. In the 
context of this problem, a flaring
magnetogram is considered as a positive while a non-flaring
magnetogram is considered a negative.
For an imbalanced dataset like this, the standard accuracy metric is
not very useful: a simple model that always predicted ``no-flare"
would have a high accuracy of $98.7\%$.  The True Skill
Statistic (TSS) score addresses this, striking an explicit balance
between correctly forecasting the positive and negative samples in a
highly-imbalanced dataset.  TSS scores range from $[-1,1]$, where a
score of 0 indicates the model doing as well as an ``always no-flare"
forecast or a chance-based forecast.  The Heidke Skill Score (HSS) is another
normalized metric used in this literature that takes values in the range 
of $[-\infty, 1]$ and reports a score of 0 for a chance-based forecast. 
Frequency bias (FB) measures the degree of overforecasting
($FB > 1$) or underforecasting ($FB<1$) in the model.

The results of these evaluation experiments, which are summarized in
Table \ref{tab:results}, show that the geometry features do almost as
well as, or slightly better than, the SHARPs features, whereas the
topology features outperform the SHARPs features by a significant
margin, as assessed by the TSS score ($\approx$ 0.05). Combining the
shape-based features with the physics-based features reveals some
useful synergies: all of the pairwise-combined feature sets outperform
the individual feature sets.  The size of the improvement varies: the
effect is somewhat stronger when geometry-based features are involved.
Interestingly, combining all three feature sets does slightly worse
than the SHARPs-topology combination: that is, simply using more
features does not guarantee better performance, a trend that has been
noted in the flare-forecasting literature, e.g. \citet{Jonas:2018}.
These improvement trends are visible across all of the metrics in the
table.

To summarize: \textit{the shape-based features outperform and/or
supplement the predictive power of the SHARPs features.}  In the
context of our MLP model, this is a particularly striking result:
abstract shape-based features automatically extracted from the
magnetic field of an active region do as well or even better than
handcrafted features viewed by experts as relevant to the physics of
an active region and the flaring process.

A look at the other metrics in Table \ref{tab:results} shows that
tuning the model for the TSS can impact some of the other metrics.  A
value of $FB > 1$---i.e., low scores for precision and high scores for
recall---indicates a high percentage of false positives (FP) and a low
percentage of false negatives (FN).  That is, our model is essentially
an overforecasting model: it sacrifices false alarms (FP) in order to
lower missed events (FN).  This is a trend observed in other
flare-prediction models in the literature, such as
DeepFlareNet \citep{Nishizuka:2018}.  Via further investigation, we
found that this is the consequence of tuning the binary cross-entropy
loss function weight.  As a consequence of tuning for the TSS metric,
this parameter takes on high values ($>$ 150), causing the model to
err on the side of correctly forecasting the flaring magnetograms.
With our hyperparameter tuning framework, it is possible to optimize
for some other metric based on the priorities of the forecaster.

\section{\revision{Deployment}}

\revision{Deployment is a major aim for us, since this research 
is proceeding in the Space Weather Technology Research and Education
Center, an organization that has a strong focus on transitioning
research models and tools to operations.  Both NOAA’s Space Weather
Prediction Center (a division of the National Weather Service) and
NASA’s Community Coordinated Modeling Center have capabilities for
comparative validation of various space weather forecasting tools.  We
will submit our final model for comparison against other solar flare
forecasting systems to one or both of these government organizations
for comparative validation. As in terrestrial weather forecasting, it
is ultimately up to the National Weather Service which tools they
choose to deploy, and those judgments are based not only on
quantitative metric comparisons but on ease of use in their
human-in-the-loop operational forecasting environment.  We are also in
discussions with the UK Met Office for evaluation and deployment of
several forecasting innovations including this solar flare prediction
model.}

\revision{As an initial step for deployment, we compared our model
with the operational flare-forecasting models evaluated
in \citet{leka2019a}.  We used a dataset similar to the one used in
that paper (training set: 2010-2015, testing set: 2016-2017), trained our
shape-based model using topological and SHARPs feature sets, and limited 
our comparison
to the M1.0+/24hr flare forecasting problem (see the top panel of
Figure 5, \citeauthor{leka2019a}, \citeyear{leka2019a}). When tuned on
the TSS metric, our proposed shape-based model returns a TSS score of 0.78, 
outperforming all the existing operational systems (TSS = [0-0.5]). However, 
our model produces a high FB score of 20.62 
(i.e., overforecasting), and performs poorly on other metrics such as 
accuracy (0.89). In comparison, the existing forecasting systems
report an FB score in the range of [0-1.5] and an accuracy of approximately 
0.95 (excluding a single outlier). Optimizing our shape-based model on the 
precision metric, on the other hand, reduces the false positives to 0, improving 
the accuracy (0.995) and FB (0.30) and making them on par with or better than
the operational forecasting models. This comes at the cost of
a lowered TSS score (0.30).}

\section{Conclusions} In this work, we introduced novel shape-based 
features constructed using tools from computational geometry and
computational topology. We successfully demonstrated their higher
forecasting capability when compared to the physics-based features
that are traditionally used in the context of a multi-layer perceptron
model. This is an important result for ML-based solar flare
forecasting research, and a stronger result than many other feature
comparison approaches---for example \citet{Chen2019}, which showed
that CNN autoencoder-extracted features from magnetograms did as well
as SHARPs-based features.

Our future directions will focus on alternative modeling approaches,
improved feature engineering, and metric optimization strategies. More
specifically, this will include validating our results with
alternative ML models (LSTMs, SVMs), improved
featurization/vectorization of persistence diagrams, performing
multivariate feature ranking to understand feature relevance with
solar flares and finally, investigating optimization trade-offs over
the different metrics using our hyperparameter tuning framework. The
feature engineering methodology in this work will eventually be
integrated into a hybrid solar flare forecasting model that will use
CNN-extracted features from solar magnetic and atmopsheric data in
combination with the physics- and shape-based features.

\section{Acknowledgements}
This material is based upon work sponsored by the National Science
Foundation Award (Grant No. AGS 2001670) and the NASA 
Space Weather Science Applications Program Award 
(Grant No. 80NSSC20K1404).

\bibliography{references}

\end{document}